\documentstyle[12pt]{article}
\begin{document}
\def\ds{\displaystyle}
\def\beq{\begin{equation}}
\def\eeq{\end{equation}}
\def\bea{\begin{eqnarray}}
\def\eea{\end{eqnarray}}
\def\ve{\vert}
\def\vel{\left|}
\def\ver{\right|}
\def\nnb{\nonumber}
\def\ga{\left(}
\def\dr{\right)}
\def\aga{\left\{}
\def\adr{\right\}}
\def\lla{\left<}
\def\rra{\right>}
\def\rar{\rightarrow}
\def\nnb{\nonumber}
\def\la{\langle}
\def\ra{\rangle}
\def\ba{\begin{array}}
\def\ea{\end{array}}
\def\tr{\mbox{Tr}}
\def\ssp{{\Sigma^{*+}}}
\def\sso{{\Sigma^{*0}}}
\def\ssm{{\Sigma^{*-}}}
\def\xis0{{\Xi^{*0}}}
\def\xism{{\Xi^{*-}}}
\def\qs{\la \bar s s \ra}
\def\qu{\la \bar u u \ra}
\def\qd{\la \bar d d \ra}
\def\qq{\la \bar q q \ra}
\def\gGgG{\la g^2 G^2 \ra}
\def\q{\gamma_5 \not\!q}
\def\x{\gamma_5 \not\!x}
\def\g5{\gamma_5}
\def\sb{S_Q^{cf}}
\def\sd{S_d^{be}}
\def\su{S_u^{ad}}
\def\ss{S_s^{??}}
\def\sbp{{S}_Q^{'cf}}
\def\sdp{{S}_d^{'be}}
\def\sup{{S}_u^{'ad}}
\def\ssp{{S}_s^{'??}}
\def\sig{\sigma_{\mu \nu} \gamma_5 p^\mu q^\nu}
\def\fo{f_0(\frac{s_0}{M^2})}
\def\ffi{f_1(\frac{s_0}{M^2})}
\def\fii{f_2(\frac{s_0}{M^2})}
\def\O{{\cal O}}
\def\sl{{\Sigma^0 \Lambda}}
\def\es{\!\!\! &=& \!\!\!}
\def\ar{&+& \!\!\!}
\def\ek{&-& \!\!\!}
\def\cp{&\times& \!\!\!}
\def\se{\!\!\! &\simeq& \!\!\!}
\def\sm{\!\!\! &\sim& \!\!\!}
\title{
         {\Large
                 {\bf
$g_{\rho\sigma\gamma}$ coupling constant
in light cone QCD
                 }
         }
      }

\author{\vspace{1cm}\\
{\small T. M. Aliev \thanks
{e-mail: taliev@metu.edu.tr}\,\,,
A. \"{O}zpineci \thanks
{e-mail: altugoz@metu.edu.tr}\,\,,
M. Savc{\i} \thanks
{e-mail: savci@metu.edu.tr}} \\
{\small Physics Department, Middle East Technical University} \\
{\small 06531 Ankara, Turkey} }
\date{}

\begin{titlepage}
\maketitle
\thispagestyle{empty}

\begin{abstract}
The coupling constant $g_{\rho\sigma\gamma}$ is determined from light cone
QCD sum rules. A comparison of our result with the ones existing in
literature is presented.  
\end{abstract}

~~~PACS number(s): 11.55.Hx, 13.40.Hq, 14.40.Cs
\end{titlepage}

\section{Introduction}
The quark model describes hadrons successfully. In its simplest version mesons
are interpreted as pure $\bar q q$ states. Scalar mesons might constitute an
exception to this successful scheme and indeed their nature
is not well established yet \cite{R1}.
At present there goes 
great deal of investigation about the identity and interpretation of the
lowest lying scalar resonances. One remarkable feature of these particles is
that they appear to be rather wide \cite{R2}--\cite{R5}. Other peculiar
property of these particles is the observation that they have very short
lifetimes and large couplings to the hadronic channels such as $K\bar K$ or
$\pi \pi$. It should be noted here that the mass of the scalar mesons varies
in the region $(400-1200)~MeV$ \cite{R6}.

Photoproduction of the $\rho^0$ meson with proton near threshold can
successfully be described by a simple one meson exchange, more precisely, by
exchange of a scalar $\sigma$ meson. An analysis of experimental data
requires to know the $g_{\rho\sigma\gamma}$ coupling constant. This coupling
constant is calculated in frame work of the effective Hamiltonian approach
and 3--point sum rules in \cite{R7,R8}, which predict
$g_{\rho\sigma\gamma}=2.71$ and $g_{\rho\sigma\gamma}=3.2 \pm 0.6$ at
$m_\sigma=0.5~GeV$,
respectively. Since this coupling constant is one of the important
quantities in analyzing the hadronic processes involving $\sigma$ meson, its
calculation in frame work of the other approaches is needed in order to get
a reliable conclusion. 

In the present note we study the coupling constant $g_{\rho\sigma\gamma}$ in
frame work of the light cone QCD sum rules (more about QCD sum rules and its
applications can be found in \cite{R9,R10}). In further analysis we assume
that the content of $\sigma$ meson is a pure $\bar q q$ state (although this
point is still under debate, see for example \cite{R11,R12}). 

The coupling of the $\sigma$ meson to the scalar current $J^\sigma =
\frac{1}{2} (\bar u u + \bar d d)$ can be parametrized in terms of a
constant $\lambda_\sigma$:
\bea
\label{cur}
\lla 0 \vel J^\sigma \ver \sigma \rra = \lambda^\sigma~.
\eea
In order to determine the coupling constant $g_{\rho\sigma\gamma}$ in frame
work of the QCD sum rules, we consider the following two point correlator
function
\bea
\label{cor1}
\Pi_\mu = i \int d^4x \, e^{i p_2 x} \lla 0 \vel T \Big\{ J^\sigma(x) 
J^\rho_\mu(0) \Big\} \ver 0 \rra_\gamma~,
\eea
where $\gamma$ means external electromagnetic field, and $J^\rho_\mu$ and
$J^\sigma$ are the interpolating currents with $\rho$ and $\sigma$ meson
numbers. The physical part of the sum rules can be obtained by inserting a
complete set of one meson states into the correlator:
\bea
\label{cor2}  
\Pi_\mu = \sum \frac{\lla 0 \vel J^\sigma(x) \ver \sigma(p_2) \rra}{p_2^2-m_\sigma^2}
\lla \sigma(p_2) \ve \rho(p_1) \rra_\gamma 
\frac{\lla \rho(p_1) \vel J^\rho_\mu(0) \ver 0 \rra}{p_1^2-m_\rho^2}~,
\eea
where $p_2=p_1+q$ and $q$ is the photon momentum. The matrix element 
$\lla \rho(p_1) \vel J^\rho_\mu(0) \ver 0 \rra$ in Eq. (\ref{cor2}) is defined as
\bea
\label{mat1}
\lla 0 \vel J^\rho_\mu \ver \rho \rra = m_\rho f_\rho \varepsilon_\mu^\rho~,
\eea
where $\varepsilon^\rho$ is the $\rho$ meson polarization vector.  
In general, the $\lla \sigma \ve \rho \rra_\gamma$ matrix can be
parametrized as  
\bea
\label{mat2}
\lla \sigma(p_2) \ve \rho(p_1) \rra_\gamma = e \Big\{ F_1(q^2) (p_1 q)
\varepsilon_\mu^\rho + F_2(q^2) (\varepsilon^\rho q) p_{1\mu} 
\Big\} \varepsilon^{\mu}~, \nnb
\eea
where $\varepsilon$ is the photon polarization vector.
In this parametrization of the matrix element $\lla \sigma\ve
\rho\rra_\gamma$, we neglect the terms $\sim q^\mu$ since 
$q^\mu \varepsilon_\mu=0$.
From gauge invariance we have 
\bea
\label{gau}
q^\mu \Big\{  F_1 (p_1 q) \varepsilon_\mu^\rho + F_2  (\varepsilon^\rho q)
p_{1\mu} \Big\} = 0 ~.
\eea
Since photon is real in our case, we immediately get from Eq. (\ref{gau})
\bea
F_2(0) = - F_1(0)~.\nnb
\eea
So, the matrix element $\lla \sigma \ve \rho \rra_\gamma$ takes the
following form:
\bea
\label{mat3}
\lla \sigma \ve \rho \rra_\gamma = e F(0) \Big\{ (p_1 q)
\varepsilon^\rho_\mu - (q \varepsilon^\rho) p_{1\mu} \Big\}
\varepsilon^\mu~.
\eea
We can use an alternative parametrization for the $\rho\sigma\gamma$ vertex,
i.e., 
\bea
\label{ant}
{\cal L}_{int} = 
\frac{e}{m_\rho} g_{\rho\sigma\gamma} 
\partial^\alpha \rho^\beta \ga \partial_\alpha A_\beta 
-\partial_\beta A_\alpha \dr ~.
\eea
When we compare Eqs. (\ref{mat3}) and (\ref{ant}) we see that
\bea
\label{com}
F(0) = \frac{g_{\rho\sigma\gamma}}{m_\rho}~.
\eea
Using Eqs. (\ref{cur}), (\ref{cor2}), (\ref{mat1}) and (\ref{mat3}),
for the phenomenological part of the sum rules we get
\bea
\label{phe}
\Pi^{phen}_\mu = g_{\rho\sigma\gamma}
\frac{\lambda_\sigma}{p_2^2-m_\sigma^2} \, 
\frac{f_\rho \varepsilon^{\nu}}{p_1^2-m_\rho^2} \Big[-(p_1 q)
g_{\mu\nu} + p_{1\nu} q_\mu  \Big]~.
\eea
Our next problem is the calculation of correlator function from QCD side.
First of all we note that the structure of the correlator is such that
perturbation contribution to it is absent due to the odd number of $\gamma$
matrix. So, QCD part of the correlator contains only nonperturbative
contribution.
Calculations lead to the following result for the theoretical part of the
correlator:
\bea
\label{bun}
\Pi^{theor}_\mu =
\frac{e \qq}{4 \pi^2} (e_u-e_d) \Big[(qx) \varepsilon_\mu -
(\varepsilon x) q_\mu \Big] \frac{\chi \varphi + g_1 x^2}{x^4}~,
\eea
where $\varphi$ and $g_1$ twist--2 and twist--4 photon wave functions,
respectively,
defined as \cite{R13}--\cite{R15}:
\bea
\label{pwf}
\la \gamma (q) \ve \bar q \sigma_{\alpha \beta} q \ve 0 \ra \es 
i e_q \qq \int_0^1 du \, e^{i u q x} \Bigg\{ (\varepsilon_\alpha q_\beta -
\varepsilon_\beta q_\alpha) \Big[ \chi \phi(u) + x^2 \Big(g_1(u) -
g_2(u)\Big)
\Big]  \nnb \\
\ar \Big[ qx (\varepsilon_\alpha x_\beta - \varepsilon_\beta x_\alpha) +
\varepsilon x (x_\alpha q_\beta - x_\beta q_\alpha) \Big] g_2
(u) \Bigg\}~,
\eea
where $\chi$ is the magnetic susceptibility of the quark condensate, $e_q$
is the quark charge.

The sum rules is obtained by equating the
phenomenological and theoretical parts (in Eq. (\ref{bun})
it is necessary to perform Fourier transformation first) of the correlator. 
Performing double Borel transformations
on the variables $p_2^2=p^2$ and $p_1^2=(p+q)^2$ on both sides of the   
correlator in order to suppress the contributions of the higher states and
the continuum (see \cite{R16}--\cite{R18}), we get the
following sum rules for the $g_{\rho\sigma\gamma}$ coupling constant:
\bea
\label{sum}
g_{\rho\sigma\gamma} e^{-\ga m_\rho^2/M_1^2 + m_\sigma^2/M_2^2 \dr} =
\frac{(e_u-e_d) \qq}{f_\rho \lambda_\sigma}
\Big[ \chi \varphi(u_0) M^2 E_0(s_0/M^2) -4 g_1(u_0) \Big]~,
\eea
where
\bea
E_0(x) = 1 - e^{-x}~,\nnb
\eea
is the function used to subtract continuum, $s_0$ is the continuum threshold
and
\bea
u_0 = \frac{M_2^2}{M_1^2+M_2^2}~,~~~~~~M^2=\frac{M_1^2 M_2^2}{M_1^2+M_2^2}~,
\nnb
\eea
where $M_1^2$ and $M_2^2$ are the Borel parameters in $\rho$ and $\sigma$
channels. Since we assume that $m_\sigma \approx 700~MeV$, which is close to
the $\rho$ meson mass, we will set $M_1^2=M_2^2=2 M^2$ and hence $u_0=1/2$.

It follows from Eq. (\ref{pwf}) that in order to determine the coupling
constant given in Eq.(\ref{sum}), one needs to know the parameters
$f_\rho$ and $\lambda_\sigma$. Leptonic decay constant of the $\rho$ meson 
is known to have the value
$f_\rho=0.18~GeV$ from experimental result of the 
$\rho \rar e^+ e^-$ decay width \cite{R19}. 
The residue $\lambda_\sigma$ is determined from sum rules obtained in \cite{R12}

\bea
\label{lmb}
\lefteqn{
\lambda_\sigma^2 e^{-m_\sigma^2/M^2} =
\frac{3 M^4}{16 \pi^2} \left\{ \left[ 1 - \ga 1 + \frac{s_0}{M^2}\dr
e^{-s_0/M^2} \right] \left[ 1 + \left( \frac{\alpha_s}{\pi} \right)
\frac{17}{3} + \left( \frac{ \alpha_s}{\pi} \right)^2 31.864 \right]\right.}
\nnb \\
\ek \left.\left( \frac{\alpha_s}{\pi} \right) \left[ 2 + \frac{95}{3}
\left( \frac{\alpha_s}{\pi} \right) \right] 
\int_0^{s_0/M^2} x \ln (x) e^{-x} dx
+ \left( \frac{ \alpha_s}{\pi}\right)^2 \frac{17}{4} \int_0^{s_0
/M^2} x \left[ \ln (x) \right]^2 e^{-x} dx \right\} \nnb \\
\ar \frac{3}{2} \lla m_q \bar{q}q\rra +
\frac{1}{16 \pi} \lla \alpha_s G^2\rra
- \frac{88 \pi}{27} \, \frac{\lla \alpha_s (\bar{q}q)^2 \rra}{M^2} \nnb \\
\ek \frac{3 \rho^2 M^6}{16 \pi^2}
e^{-\rho^2 M^2/ 2} \left[ K_0 \left( \frac{\tau^2 M^2}{2}
\right) + K_1 \left( \frac{\tau^2 M^2}{2} \right) \right],
\eea
where the last term describes the single instanton contribution, $K_i$ are
the modified Bessel functions and $\tau$
is the instanton size for which we will use $\tau=(0.6~GeV)^{-1}$ (see
\cite{R20}).

In further numerical analysis we use the following values for the input
parameters: $\lla m_q \bar{q}q \rra = (-0.82 \pm 0.1)\times 10^{-4}~GeV^4$,
$\lla \alpha_s G^2 \rra = (0.038 \pm 0.011)~GeV^4$, 
$\lla \alpha_s (\bar{q}q)^2 \rra = (-0.18 \pm 0.1)\times 10^{-3}~GeV^4$
 (see for example \cite{R10}). For the mass of the scalar meson we use
$m_\sigma \approx 700~MeV$. The dependence of $\lambda_\sigma$ on Borel
parameter $M^2$ at three values of the continuum threshold $s_0=(1.4;~1.6;$
and $1.8)~GeV^2$ is presented in Fig. (1). Since the Borel parameter $M^2$
has no physical meaning, we must look for the so--called stability region 
where sum rules is practically independent of $M^2$. We observe from 
Fig. (1) that the stability window $M^2$ lies in the interval $1.2~GeV^2 <
M^2 < 1.4~GeV^2$ and we get
\bea
\label{rlm}
\lambda_\sigma = (0.2 \pm 0.02)~GeV^2~,
\eea
where the error can be attributed to the variation of the continuum
threshold $s_0$, Borel parameter $M^2$ and the errors in the condensates
$\qq$ and $\lla G^2 \rra$. It should be noted that if we had used the values 
of the input parameters given in \cite{R8} we get 
\bea
\label{iff}
\lambda_\sigma = (0.16 \pm 0.02)~GeV^2~. \nnb
\eea
This difference arises from the omitted
instanton contribution and terms in perturbative contribution in \cite{R8}.

Having the values of $\lambda_\sigma$ and $f_\rho$, 
our next goal is calculating the coupling constant
$g_{\rho\sigma\gamma}$ from Eq. ({\ref{sum}).
As is obvious from Eq. (\ref{sum}) the main input parameters of
the sum rule is photon wave functions. We shall make use of the following
expressions for the photon wave functions \cite{R13,R15}
\bea
\varphi(u) \es 6 u (1-u)~, \nnb \\
g_1(u) \es -\frac{1}{8} (1-u) (3-u)~,\nnb
\eea
and for the magnetic susceptibility we use $\chi = - 4.4~GeV^{-2}$
\cite{R21}.

In Fig. (2) we present the dependence of the coupling constant
$g_{\rho\sigma\gamma}$ on the Borel parameter $M^2$at three different values
of the continuum threshold: $s_0=1.4~GeV^2;~1.6~GeV^2;~1.8~GeV^2$. It follows
from this figure that for the choices $s_0=1.4~GeV^2$ and $s_0=1.8~GeV^2$
the variation in the result is about $\sim 10\%$, i.e., the coupling
constant can be said to be practically independent of the continuum
threshold $s_0$. Furthermore the coupling constant seems to be insensitive
to the variation of the Borel parameter $M^2$. Along these lines,
we calculated the coupling constant $g_{\rho\sigma\gamma}$ and our final
result is
value
\bea
\label{cou}
g_{\rho\sigma\gamma} = (2.2 \pm 0.4)~,
\eea
where all possible sources of uncertainties are taken into account, namely,
errors coming from determination of $\lambda_\sigma$, from variation of the
continuum threshold $s_0$, Borel parameters $M^2$, neglected twist--3 
photon wave functions and errors in the values of the condensates.   
   
Finally, we would like to present a comparison of our prediction on 
the coupling constant $g_{\rho\sigma\gamma}$ with the
existing results in literature. Traditional 3--point QCD sum rules analysis
predicts $g_{\rho\sigma\gamma} = (3.2 \pm 0.6)$ \cite{R8}. Our result is
about $50\%$ lower than this result, which can be attributed to the
difference in the values of $\lambda_\sigma$. More over our result is
slightly lower than the result obtained from effective Lagrangian approach,
which predicts $g_{\rho\sigma\gamma} = 2.71$ \cite{R7}.

\newpage

\section*{Figure captions}
{\bf Fig. (1)} The dependence of $\lambda_\sigma$ on Borel parameter
$M^2$ at three different values of the continuum threshold $s_0=1.4~GeV^2;
1.6 ~GeV^2$ and $s_0=1.8~GeV^2$. \\ \\
{\bf Fig. (2)} The dependence of the $g_{\rho\sigma\gamma}$ coupling constant
on Borel parameter $M^2$ at three different values of the continuum
threshold $s_0=1.4~GeV^2;   
1.6 ~GeV^2$ and $s_0=1.8~GeV^2$.

\newpage

\begin{figure}
\vskip 1cm
    \includegraphics{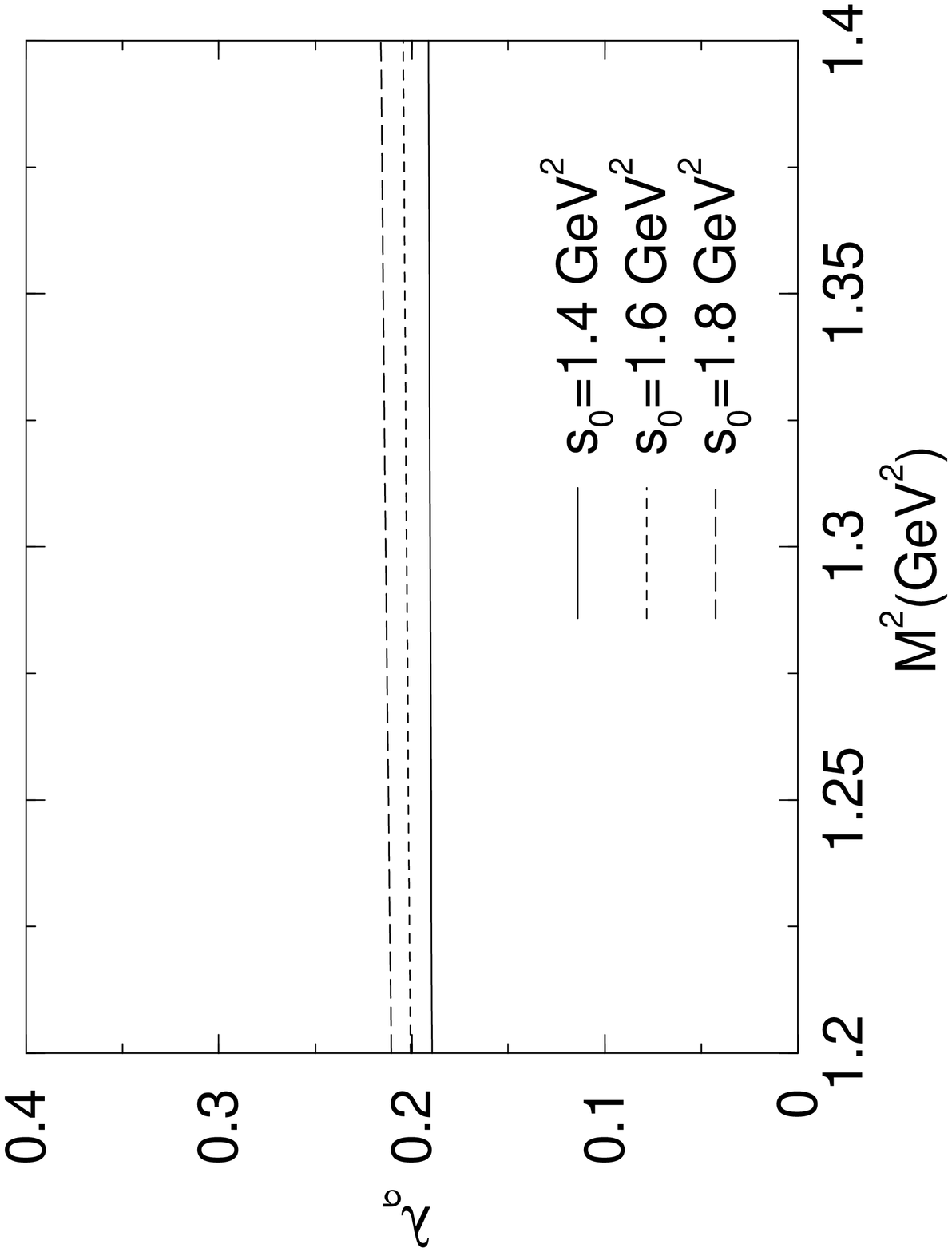}
\vskip 9.cm
\begin{center}
Figure {\bf 1}  
\end{center}
\end{figure}

\begin{figure}  
\vskip 1.5 cm
    \includegraphics{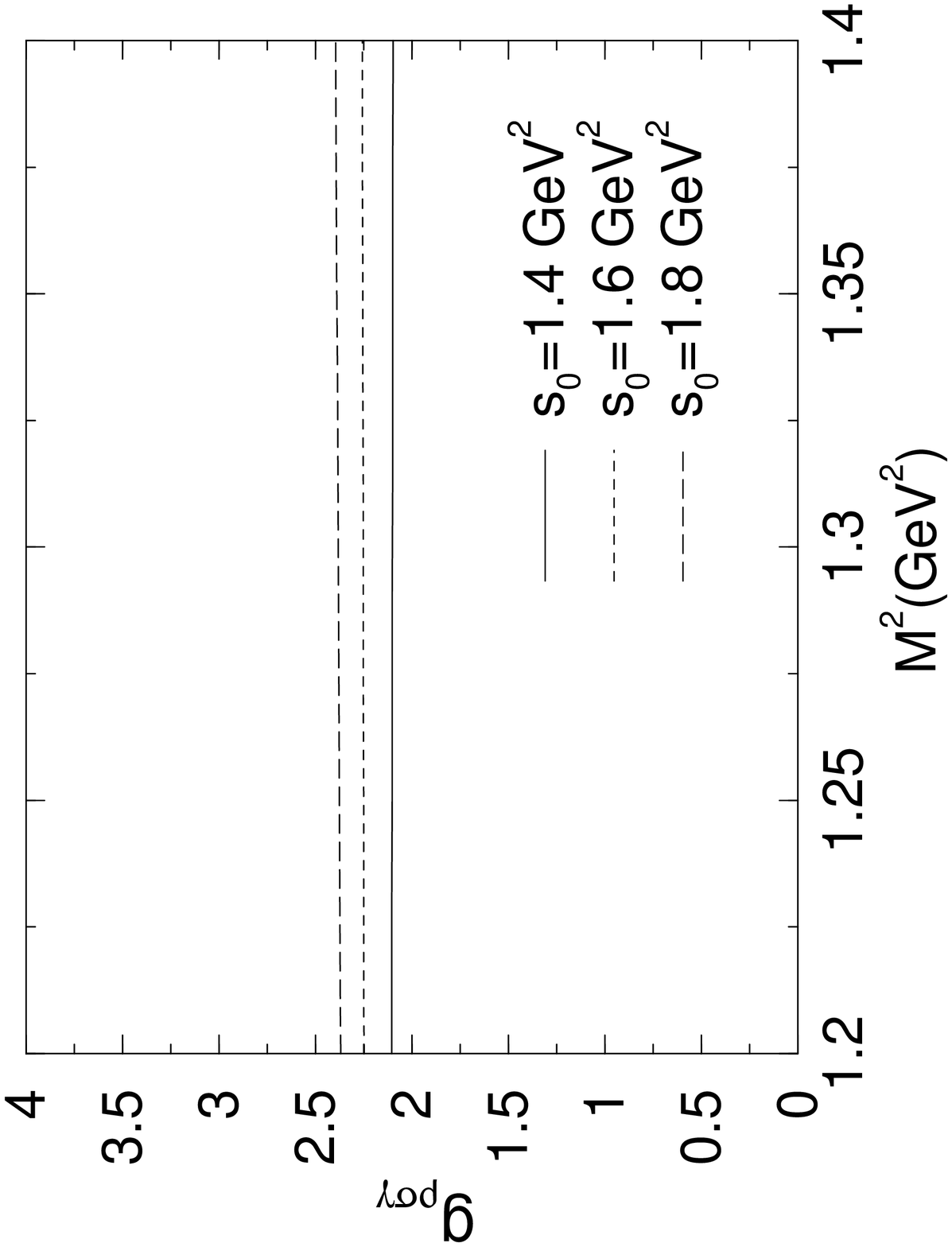}
\vskip 9. cm
\begin{center}   
Figure {\bf 2}   
\end{center}
\end{figure}

\end{document}